\documentclass{PoS}
\usepackage{braket}
\renewcommand{\d}{\ensuremath{\mathrm{d}}}

\newcommand{\gf}{\ensuremath{\mathrm{gf}}}

\newcommand{\YM}{\ensuremath{\mathrm{YM}}}

\newcommand{\p}{\partial}

\newcommand{\RGZ}{\ensuremath{\mathrm{RGZ}}}
\newcommand{\GZ}{\ensuremath{\mathrm{GZ}}}
\newcommand{\CGZ}{\ensuremath{\mathrm{CGZ}}}

\newcommand{\FP}{\ensuremath{\mathrm{FP}}}

\title{More evidence for a refined Gribov-Zwanziger action based on an effective potential approach}

\ShortTitle{More evidence for a RGZ action based on an effective potential approach}

\author{\speaker{Nele Vandersickel}, David Dudal \\
        Ghent University, Department of Physics and Astronomy \\
        Krijgslaan 281-S9, 9000 Gent,Belgium\\
        E-mail: \email{nele.vandersickel@ugent.be} ,\email{ david.dudal@ugent.be},}

\author{Silvio P. Sorella\\
        Departamento de F\'{\i }sica Te\'{o}rica, Instituto de F\'{\i }sica, UERJ - Universidade do Estado do Rio de Janeiro\\
        Rua S\~{a}o Francisco Xavier 524, 20550-013 Maracan\~{a}, Rio de Janeiro, Brasil\\
        E-mail: \email{sorella@uerj.br}
}

\abstract{The purpose of this proceeding is twofold. Firstly, we shall make the refining of the Gribov-Zwanziger action more complete by taking into account more condensates than considered so far. Secondly, we shall provide more evidence for the refined Gribov-Zwanziger action based on an effective potential approach.}

\FullConference{The many faces of QCD\\
		November 1-5, 2010\\
		Gent Belgium}

\begin{document}

\section{Introduction}
\noindent The infrared behavior of propagators has received a lot of interest in recent years. Many of the discussions have evolved around the zero momentum value of the gluon propagator and the enhancement of the ghost. Until recently, the lattice calculations and analytical approaches agreed on a vanishing gluon propagator at zero momentum and a ghost propagator which is infrared enhanced. One particular model, the Gribov-Zwanziger framework, which takes into account Gribov copies, also agreed on this behavior. However, in the last couple of years, the understanding of the behavior of the propagators has changed dramatically. The common believe is now that the ghost propagator is not enhanced and the gluon propagator attains a finite nonvanishing value at zero momentum. Such propagators were found by lattice as well as by many analytical approaches \cite{Cucchieri:2007md,Bogolubsky:2007ud,Aguilar:2008xm,Fischer:2008uz,Pene:2009iq,Dudal:2010tf}. In particular, in the Gribov-Zwanziger framework, the new behavior of the ghost and gluon propagator was explained by taking into account the existence of a certain dimension 2 condensate. This framework was called the refined Gribov-Zwanziger framework \cite{Dudal:2007cw,Dudal:2008sp}.\\
\\
The condensate of interest in the refined GZ action  stems from a BRST invariant  operator of dimension $d=2$. However, as is well known, the GZ action breaks BRST symmetry,  so there is no reason to a priori investigate only BRST invariant operators.  In fact, more $d=2$ condensates can be investigated without spoiling the renormalizability. These condensates are dynamical objects, giving rise to dynamical mass scales. To get an estimate for the latter, we have to compose the effective potential and look for the minimum of this potential. We shall do this for the first time,  our aim being that of providing  evidence that these condensates are in fact present.

\section{The Gribov-Zwanziger action}
\noindent Let us start by reviewing the Gribov-Zwanziger action. The Yang-Mills action is given by
\begin{eqnarray}
S_\YM &=&  \frac{1}{4}\int \d^4 x\; F_{\mu\nu}^a F_{\mu\nu}^a \;.
\end{eqnarray}
However,  the Yang-Mills action needs to be gauge fixed in order to define the path integral, as else, it is ill-defined. This can be done with the Faddeev-Popov action  which, in the Landau gauge,  $\p_\mu A_\mu = 0$, reads
\begin{equation}
S_\FP = S_\YM + S_\gf \;,
\end{equation}
with
\begin{equation}
S_\gf = \int \d^4 x \left( b^{a}\partial_\mu A_\mu^{a}+\overline{c}^{a}\partial _{\mu } D_{\mu}^{ab}c^b \right) \;,
\end{equation}
being the gauge fixing part. This action enjoys BRST symmetry, given by
\begin{equation}\label{s}
    sA_{\mu }^{a} =-\left( D_{\mu }c\right) ^{a}\;,    \hspace{0.5cm}
    sc^{a} =\frac{1}{2}gf^{abc}c^{b}c^{c}\;, \hspace{0.5cm}
    s\overline{c}^{a} =b^{a}\;, \hspace{0.5cm}
    sb^{a}=0\;, \hspace{0.5cm} s^2 = 0\;,
\end{equation}
which is important to prove perturbative unitarity and renormalizability. However, as shown by Gribov, this gauge fixing suffers from Gribov copies \cite{Gribov:1977wm}. If we take e.g.~the Landau gauge, $\partial_\mu A_\mu = 0$, it is very easy to show that there exist gauge equivalent fields $A'_\mu$ also fulfilling the Landau gauge $\p_\mu A_\mu' = 0$ when having zero modes of the Faddeev-Popov operator
\begin{eqnarray}
 \mathcal M^{ab} &=&  - \p_\mu D_\mu^{ab} = -\p_\mu (\p_\mu \delta^{ab} - g f^{abc} A_\mu^c)\;.
\end{eqnarray}
Therefore, Gribov proposed to restrict the region of integration  in the path integral to $\Omega$,
\begin{eqnarray}
\Omega=\left\{  A_{\mu}|\partial_{\mu} A_{\mu}=0, \mathcal M>0  )    \right\}\;,
\end{eqnarray}
the region which is free of zero modes of the Faddeev-Popov operator $\mathcal M^{ab}$. Unfortunately, it is not so difficult to show that there are still Gribov copies inside the Gribov region, as first proven in \cite{Semenov}.  In 1989, Zwanziger implemented this restriction to the Gribov region to all orders: the Gribov-Zwanziger action was born \cite{Zwanziger:1989mf}. We present immediately the localized version of this action, namely
\begin{equation}
S_\GZ = S_{\FP} + S_0+ S_{\gamma}\;,
\end{equation}
with $S_{\FP}$ the usual Faddeev-Popov action,
\begin{equation}
S_{\FP} = \frac{1}{4}\int \d^4 x F_{\mu\nu}^a F_{\mu\nu}^a +  \int \d^4 x \left( b^{a}\partial_\mu A_\mu^{a}+\overline{c}^{a}\partial _{\mu } D_{\mu}^{ab}c^b \right) \;,
\end{equation}
and $S_0$ and $S_\gamma$ given by
\begin{eqnarray}
S_0 &=& \int \d^{4}x\left( \overline{\varphi}_\mu^{ac} \mathcal M^{ab}\varphi^{bc}_\mu -\overline{\omega}_{\mu}^{ac} \mathcal M^{ab} \omega_\mu^{bc} \right) \;, \nonumber\\
S_{\gamma}&=& -\gamma ^{2}g\int\d^{4}x\left( f^{abc}(\varphi _{\mu }^{bc} +\overline{\varphi }_{\mu }^{bc} )A_{\mu }^{a} + \frac{d}{g}\left(N^{2}-1\right) \gamma^{2} \right) \;. \nonumber
\end{eqnarray}
The fields $( \overline{\varphi }_{\mu}^{ac},\varphi_{\mu}^{ac}) $ are a pair of complex conjugate bosonic fields, while $( \overline{\omega }_{\mu}^{ac},\omega_{\mu}^{ac}) $ are anticommuting fields. $\gamma$ is not free, but fixed by the horizon condition,
\begin{equation}\label{gapgamma}
\frac{\p \Gamma}{\p \gamma} = 0\;,
\end{equation}
whereby $\Gamma$ is the quantum action defined as,
\begin{equation}
e^{-\Gamma} = \int [\d \phi] e^{-S}\;,
\end{equation}
and $[\d \phi]$ represents the integration over all the fields. In the case of the local action, this condition is equivalent to,
\begin{equation}\label{horizon}
\Braket{ gf_{cka}A_{\mu}^{k}( \varphi_{\mu}^{ac}+\overline{\varphi}_{\mu}^{ac} ) } = 2\gamma^{2}4(N^{2}-1)\;.
\end{equation}
Looking at the BRST transformation of the new fields\footnote{We shall use a shorthand notation for the new fields: e.g.~$\varphi_\mu^{ac} = \varphi_i^a$. },
\begin{equation}
s\varphi _{i}^{a} =\omega _{i}^{a}\;, \hspace{0.5cm} s\omega _{i}^{a}=0\;, \hspace{0.5cm} s\overline{\omega }_{i}^{a} =\overline{\varphi }_{i}^{a}\;,\hspace{0.5cm} s \overline{\varphi }_{i}^{a}=0\;,
\end{equation}
one can check very easily that $S_{\GZ}$ is no longer invariant under the BRST symmetry $s$,
\begin{eqnarray}
 sS_\GZ = s \left(S_{\mathrm{YM}}+S_{gf} + S_0+ S_{\gamma}\right)  = s \left(S_\gamma \right) = g \gamma^2 \int \d^d x f^{abc} \left( A^a_{\mu} \omega^{bc}_\mu -
 \left(D_{\mu}^{am} c^m\right)\left( \overline{\varphi}^{bc}_\mu + \varphi^{bc}_{\mu}\right)  \right)\,.\nonumber\\
\end{eqnarray}
Therefore, this action breaks the famous BRST symmetry which,  among other things,  leaves open the question of unitarity. For more on this breaking, we refer to \cite{Dudal:2007cw}. Let us mention that when removing the Gribov horizon, i.e.~setting $\gamma = 0$, there is no breaking and the extra fields $(\varphi, \overline \varphi, \omega, \overline \omega)$ give rise to a BRST quartet and decouple from the theory. It is remarkable that the breaking of the BRST does not spoil the renormalizability of the GZ action, due to the soft nature of the breaking. \\
\\
Let us have a look at the ghost and  gluon propagators predicted by the Gribov-Zwanziger model \cite{Gribov:1977wm,Zwanziger:1989mf,Ford:2009ar,Gracey:2005cx}. Firstly, the gluon propagator is  positivity violating, infrared suppressed and vanishing at zero momentum. E.g.~the tree level gluon propagator is given by,
\begin{equation}\label{nice}
 \braket{A_\mu^a (p) A_\nu^b (-p) } =  \delta^{ab} \left( \delta_{\mu\nu}- \frac{p_\mu p_\nu }{p^2}\right)\underbrace{ \frac{p^2 }{p^4 + 2 g^2 N \gamma^4 } }_{D(p)} \;,
\end{equation}
whereby also at higher loops, the gluon propagator becomes zero at zero momentum \cite{Gracey:2005cx}. Secondly, the ghost propagator is enhanced at zero momentum,
\begin{eqnarray}
\lim_{k\to 0} G(k^2) \sim \frac{1}{k^4} \;,
\end{eqnarray}
which has also been explicitly checked up to two loops, \cite{Ford:2009ar,Gracey:2005cx}. However, both results do not seem to agree with the lattice results, see \cite{Cucchieri:2007md,Cucchieri:2007rg}.

\section{Refining the Gribov-Zwanziger action}
\noindent As the horizon condition (\ref{horizon}) is equivalent with a dimension two condensate, it would appear naturally that other such condensates could appear in the Gribov-Zwanziger approach. In \cite{Dudal:2007cw}, we have investigated the following two condensates, $\Braket{A^2}$ and  $\Braket{\overline \varphi^a_i \varphi^a_i - \overline \omega^a_i \omega^a_i }$, which gave rise to the refined Gribov-Zwanziger action,
\begin{equation}
S_\RGZ = S_\GZ + \int \d^4 x \left( \frac{1}{2} m^2 A_\mu^a A_\mu^a -  M^2 (\overline \varphi^a_i \varphi^a_i - \overline \omega^a_i \omega^a_i) \right)\;.
\end{equation}
The masses $m^2$ and $M^2$ associated with the condensates   have a dynamical origin.  \\
\\
The propagators of the refined action display a different behavior now. Firstly, the gluon propagator remains infrared suppressed and positivity violating, but does not vanish anymore at zero momentum. At tree level it is given by
\begin{equation}\label{nice}
D(p^2) = \frac{p^2 + M^2}{p^4 + \left(M^2 + m^2\right) p^2 + 2 g^2 N \gamma^4 + M^2 m^2  }\;,
\end{equation}
and it remains non zero at higher loop order \cite{Dudal:2008sp}.
Secondly, the ghost propagator behaves $\propto 1/k^2$ and is thus not enhanced anymore. The behavior of the propagators is in qualitative agreement with the lattice data of \cite{Cucchieri:2007md,Cucchieri:2007rg}. For some quantitative numbers, we refer to \cite{Dudal:2010tf}, whereby the tree level gluon propagator (\ref{nice}) was fitted to the lattice data. The following estimates were found,
\begin{eqnarray}
m^2 &\approx& -1.78\ \ensuremath{\mathrm{ GeV}}^2 \;, \nonumber\\
M^2 &\approx& 2.14 \  \ensuremath{\mathrm{ GeV}}^2 \;,\nonumber\\
\lambda^4 &\approx& 0.26\ \ensuremath{\mathrm{ GeV}}^4 \nonumber\\
D(0) &\approx& 8.2 \ \ensuremath{\mathrm{ GeV}}^2  \;,\nonumber\\
\braket{g^2 A^2}_{\mu = 10\ \ensuremath{\mathrm{ GeV}}} &\approx& 3\ \ensuremath{\mathrm{ GeV}}^2 \;,
\end{eqnarray}
in association with a very good quality of the fit, and with the value of $\braket{g^2 A^2}$ in the ballpark of other estimates from different approaches \cite{Boucaud:2008gn,RuizArriola:2004en} .

\section{A further refining of the GZ action}
\noindent So far, the main ingredient in the refining of the GZ action was the inclusion of a dimension 2 condensate for which the corresponding operator is BRST invariant, namely $\overline{\varphi}^a_i \varphi^a_{i} - \overline{\omega}^a_i \omega^a_i = s(\overline \omega^a_i \varphi^a_i)$. However, as we have explained, the GZ action breaks the BRST symmetry, so there is no reason to only investigate this $d=2$ BRST invariant operator. In fact, not only are there more possible condensates which can be present, we shall also split the operator $\overline{\varphi}^a_i \varphi^a_{i} - \overline{\omega}^a_i \omega^a_i$ into two separate pieces. All possible renormalizable $d=2$ operators $\mathcal O_i$ in the GZ action, which have ghost number zero, are given by
\begin{equation}\label{operators}
\mathcal O_i = \{   A^a_\mu A^a_\mu,  \varphi_i^a  \varphi_i^a, \varphi_i^a \overline \varphi_i^a,   \overline{\varphi}^a_i \overline \varphi^a_i , \overline \omega^a_i \omega^a_i \}\;.
\end{equation}
We choose to only investigate operators  for  which both color and Lorentz indices are fully contracted, e.g.~like $\varphi_i^a \overline \varphi_i^a = \varphi_\mu^{ac} \overline \varphi_\mu^{ac}$. However, as one can find in \cite{Gracey:2010cg}, there are other possibilities to combine the color indices. If one wants to be absolutely complete, one would have to take into account all possible color contractions. Unfortunately, doing this would be hopelessly complicated and we hope that we have captured the mayor physics by taking only one color combination. However, in principle, different color combinations are possible.\\
\\
We have added all the $d=2$ operators $\mathcal O_i$ to the GZ action, and we have proven that this can be done in a renormalizable fashion \cite{Nele}. As there is no lowest order coupling of $\omega$ and $\overline \omega$ to the gluon sector, for the calculation of the effective action, we shall not take this condensate into account. The enlarged action is then given by
\begin{eqnarray}\label{startxx}
\Sigma_\CGZ &=& S_\GZ + \int \d^4 x  \left[ Q  \overline \varphi^a_i \varphi^a_i   + \frac{1}{2} \tau A_{\mu }^{a}A_{\mu }^{a}  - \frac{1}{2}\zeta \tau ^{2}  - \alpha   Q Q    - \chi Q \tau\right] \nonumber\\
&& + \int \d^4 x \left[ \frac{1}{2} \overline G \overline \varphi^a_i \overline \varphi^a_i + \frac{1}{2} G \varphi^a_i \varphi^a_i  + \varrho G \overline G \right]\;,
\end{eqnarray}
whereby $Q$, $\tau$,  $G$ and $\overline G$ are the new sources coupled to the relevant operators, while $\zeta$, $\alpha$, $\chi$ and $\varrho$ are the dimensionless LCO parameters  of the quadratic terms in the sources which are needed to account for the divergences present in the correlation functions like $\braket{\mathcal O_i(k) \mathcal O_j(-k) }$, with $\mathcal O_i$ one of the operators given in expression (\ref{operators}), see \cite{Verschelde:2001ia}. This action is multiplicative renormalizable, however, the renormalization is non-trivial. Due to mixing of the sources $Q$ and $\tau$, the renormalization requires some effort \cite{Nele}.\\
\\
Let us again discuss the propagators. The gluon propagator becomes slightly more complicated, although remains infrared suppressed and non-zero at zero momentum. The tree level expression is given by
\begin{equation}
D(p^2) = \frac{ \left(M^2+p^2\right)^2-  \rho  \rho^{\dagger}  }{  M^4 p^2+  p^6+ M^2 \left(2 p^4+\lambda^4 \right)-\lambda^4  \frac{(\rho +\rho^{\dagger} )}{2}+ m^2 \left(\left(M^2+p^2\right)^2-\rho  \rho^{\dagger} \right)+ p^2 (\lambda^4 -\rho  \rho^{\dagger} )}\;.
\end{equation}
Notice that this propagator reduces to the form of the RGZ propagator (\ref{nice}) in the case that $\rho = \rho^\dagger$,
\begin{eqnarray}
D(p^2) = \frac{M^2+p^2+\rho }{p^4 + M^2 p^2 +p^2 (\rho + m^2) +m^2 \left(M^2+\rho \right) +  \lambda ^4} \;,
\end{eqnarray}
Comparison with existing lattice data shall determine whether this is the case or not. Secondly, the ghost propagator remains not enhanced, similar to the previous section.

\section{The effective action}
\noindent Fortunately, we shall be able to calculate the effective action at lowest order. For this, we shall follow the local composite operator (LCO) formalism developed  in \cite{Verschelde:2001ia,Verschelde:1995jj}. In fact, it is interesting to notice that before is was not possible to apply the LCO formalism due to the specific form of the operator $\overline \varphi^a_i \varphi^a_i - \overline \omega^a_i \omega^a_i$, as we were not able to perform a Hubbard Stratonovich transformation to find the effective action\footnote{More details on this will be provided in \cite{Nele}.}. However, by splitting this operator as we did, it is possible to apply the LCO formalism.  After a long analysis, we were able to calculate the effective action\cite{Nele},
\begin{eqnarray}\label{finaleffective}
\Gamma^{(1)} &=&  \frac{(N^2 -1)^2}{16 \pi^2} \Bigl[(M^2 - \sqrt{\rho \rho^\dagger})^2 \ln\frac{M^2 - \sqrt{\rho \rho^\dagger}}{\overline \mu^2}  + (M^2 + \sqrt{\rho \rho^\dagger})^2 \ln\frac{M^2 + \sqrt{\rho \rho^\dagger}}{\overline \mu^2} \nonumber\\
&& - 2 (M^2 + \rho \rho^\dagger)\Bigr] + \frac{3(N^2 -1)}{64\pi^2} \Bigl[  - \frac{5}{6} (m^4 -2 \lambda^4) + y_1^2 \ln \frac{(-y_1)}{\overline \mu} + y_2^2 \ln \frac{(-y_2)}{\overline \mu} + y_3^2 \ln \frac{(-y_3)}{\overline \mu} \nonumber\\
&&- y_4^2 \ln \frac{(-y_4)}{\overline \mu} - y_5^2 \ln \frac{(-y_5)}{\overline \mu} \Bigr]  -2 (N^2 - 1) \frac{\lambda^4}{N g^2}  + \frac{3}{2} \frac{\lambda^4}{32 \pi^2} (N^2-1) \nonumber\\
&&+ \frac{1}{2} \frac{48 (N^2-1)^2}{53 N} \left( 1   - N g^2 \frac{53}{24} \frac{\varrho_1}{(N^2 - 1)^2} \right) \frac{\rho \rho^\dagger}{g^2} \nonumber\\
&&  +  \frac{9}{13} \frac{N^2-1}{N}\frac{m^4}{2g^2}- \frac{24}{35}\frac{(N^2-1)^2}{N}\frac{M^4}{g^2}  - \frac{161}{52} \frac{N^2-1}{16 \pi^2}\frac{ m^4}{2 }- M^4 \alpha_1  +M^2 m^2 \chi_1\;.
\end{eqnarray}
whereby $y_1$, $y_2$ and $y_3$ are the solutions of the equation $y^3+(m^2 +2 M^2) y^2 +\bigl(\lambda^4+ M^4- \rho \rho^\dagger +2 M^2 m^2 \bigr) y + M^2 \lambda^4 + 1/2 ( \rho + \rho^\dagger) \lambda^4 + M^4 m^2  - m^2 \rho \rho^\dagger  =0$ and $y_4$ and $y_5$ of the equation $ y^2 + 2 M^2 y +M^4 -\rho \rho^\dagger =0$. It is not only necessary to calculate all the one loop diagrams of the theory, also some two loop information is needed in order to determine the one loop effective action \cite{Verschelde:2001ia,Verschelde:1995jj}. However, this remains  for future work, so we have parameterized the unknown information  into the three constants $\varrho_1$, $\alpha_1$ and $\chi_1$.\\
\\
In order for the condensates to be indeed present, we have to show that the minimum of the effective action requires a non zero value at least for one of the condensates $\braket{\overline \varphi \overline \varphi}$, $\Braket{\varphi \varphi}$ or $\Braket{\varphi \overline \varphi}$. In order to show this, let us simplify the effective action by setting $\rho= \rho^\dagger = 0$. This is equivalent with not considering the condensates $\braket{\overline \varphi \overline \varphi}$ and $\Braket{\varphi \varphi}$ and only taking  $\braket{\varphi \overline \varphi}$ into account. We remind the reader that already a non zero value of the condensate $\braket{\varphi \overline \varphi}$ is already sufficient for $D(0) \not=0$. In this case, the effective action becomes:
\begin{eqnarray}\label{mineffpot}
\Gamma^{(1)} &=&  \frac{(N^2 -1)^2}{16 \pi^2}\Bigl[2 M^4 \ln\frac{M^2}{\overline \mu^2}   - 2 M^2 \Bigr] + \frac{3(N^2 -1)}{64\pi^2} \Bigl[  - \frac{5}{6} (m^4 -2 \lambda^4)+ M^4 \ln \frac{(M^2)}{\overline \mu} \nonumber\\
&&+ y_2^2 \ln \frac{(-y_2)}{\overline \mu} + y_3^2 \ln \frac{(-y_3)}{\overline \mu} - 2M^4 \ln \frac{M^2}{\overline \mu}  \Bigr]  -2 (N^2 - 1) \frac{\lambda^4}{N g^2}  + \frac{3}{2} \frac{\lambda^4}{32 \pi^2} (N^2-1) \nonumber\\
&&  +  \frac{9}{13} \frac{N^2-1}{N}\frac{m^4}{2g^2}- \frac{24}{35}\frac{(N^2-1)^2}{N}\frac{M^4}{g^2}  - \frac{161}{52} \frac{N^2-1}{16 \pi^2}\frac{ m^4}{2 }- M^4 \alpha_1  +M^2 m^2 \chi_1\;.
\end{eqnarray}
whereby $y_2$ and $y_3$ are given by $\frac{1}{2} \left(-m^2-M^2\pm\sqrt{m^4-2 M^2 m^2 + M^4 - 4 \lambda^4}\right)$. \\
\\
In order to find the minimum, we should derive this action w.r.t.~$m^2$ and $M^2$ and put the equations equal to zero. In addition, we should also impose the horizon condition (\ref{gapgamma}). Therefore, we have the following three conditions,
\begin{equation}
\frac{\p \Gamma}{\p M^2} =0 \;, \hspace{1.5cm}  \frac{\p \Gamma}{\p m^2} =0 \;, \hspace{1.5cm}   \frac{\p \Gamma}{\p \lambda^4} =0 \;,
\end{equation}
which have to be solved for $M^2$, $m^2$ and $\lambda^4$. Unfortunately, it is impossible to solve these equations exactly due to the two unknown parameters $\alpha_1$ and $\chi_1$.  However, we can uncover if $M^2 = 0$ can be a solution of the above expression. Looking at the three conditions, it is clear that we will obtain three equations with only two unknown parameters $(m^2, \lambda^2)$. Although one of these equations still contains the unknown constant parameter $\chi_1$, it is highly unlikely that this system of equations will have a solution for $(m^2, \lambda^2)$.  We take this as an indication that the condensate $\braket{\overline \varphi \varphi}$ is indeed present, thereby suggesting the dynamical transformation of GZ into a refined GZ.

\section{Conclusion}
\noindent In this paper, we have made the refining of the GZ action more complete. We have shown that more $d=2$ condensates are possible  than the ones so far considered, i.e.~$\braket{A_\mu^a A_\mu^a}$ and $\Braket{\overline \varphi^a_i \varphi^a_i - \overline \omega^a_i \omega^a_i }$. Moreover, as the GZ action breaks the BRST symmetry anyway, there is no reason to consider the operator  $\overline \varphi^a_i \varphi^a_i - \overline \omega^a_i \omega^a_i $ as a whole. In summary, we have taken the following operators into account to further refine the GZ action:
\begin{equation}\label{operators}
\mathcal O_i = \{ A^a_\mu A^a_\mu,  \varphi_i^a  \varphi_i^a, \varphi_i^a \overline \varphi_i^a,   \overline{\varphi}^a_i \overline \varphi^a_i  \}\;,
\end{equation}
while not spoiling the renormalizability.\\
\\
A second  aim  was to calculate the one loop effective action including all the condensates. We have succeeded in finding this effective action, see expression (\ref{finaleffective}), although there are still some unknown constants stemming from two loop diagrams. Though, we are able to show that it is very likely that the minimum of the effective potential requires a non zero value of the condensate $\varphi_i^a \overline \varphi_i^a$. This would prove the existence of a refined GZ action from a dynamical point of view.

\section*{Acknowledgments}
\noindent We are grateful to D. Zwanziger for useful discussions. We would also like to thank all the participants of the workshop ``The many faces of QCD'' for interesting discussions. D.~Dudal and N.~Vandersickel are  supported by the Research Foundation-Flanders (FWO). The Conselho Nacional de  Desenvolvimento Cient\'{\i}fico e Tecnol\'{o}gico (CNPq-Brazil), the Faperj,
 Funda{\c{c}}{\~{a}}o de Amparo {\`{a}} Pesquisa do Estado  do Rio de Janeiro, the SR2-UERJ and the Coordena{\c{c}}{\~{a}}o de  Aperfei{\c{c}}oamento de Pessoal de N{\'{\i}}vel Superior (CAPES),
 the CLAF, Centro Latino-Americano de F{\'\i}sica, are gratefully acknowledged for financial support.

\end{document}